\documentclass[rnaas]{aastex631}

\shorttitle{Optical Photometry of WR140}
\shortauthors{Peatt et al.}
\graphicspath{{./}{figures/}}

\begin{document}

\title{Optical Photometry of WR140 as the Dust Formed During the 2016 Periastron Passage}

\correspondingauthor{Noel D. Richardson}
\email{noel.richardson@erau.edu}

\author{Megan J. Peatt}
\affiliation{Department of Physics and Astronomy \\ Embry-Riddle Aeronautical University \\
3700 Willow Creek Rd \\
Prescott, AZ 86301, USA}

\author[0000-0002-2806-9339]{Noel D. Richardson}
\affiliation{Department of Physics and Astronomy \\ Embry-Riddle Aeronautical University \\
3700 Willow Creek Rd \\
Prescott, AZ 86301, USA}
\begin{abstract}

The colliding wind binary WR140 produces dust in its shocked gas every periastron passage. While the infrared light curve is very repeatable, there are noticeable changes every cycle in the optical time-series photometry. In the phases following periastron, there are optical dips in the light curve that were postulated to be caused by localized clumps in the dust produced in our line of sight. We report on the $B$ and $V$-band light curves that were recorded by the American Association of Variable Star Observers (AAVSO) after the 2016 periastron event and briefly discuss comparisons to geometric models of the dust production to infer that these features are likely caused by localized dust clumps in the new dust shell.

\end{abstract}


\section{Introduction} \label{sec:intro}

Classical Wolf-Rayet stars are helium-burning, evolved massive stars that have lost their outer hydrogen envelopes. These stars have strong stellar winds with high mass-loss rates and wind speeds of several thousand km s$^{-1}$. These stars can be either nitrogen-rich (WN) or carbon-rich (WC) in their classification \citep[see review by ][]{2007ARA&A..45..177C}. Some WC stars have been shown to have very large infrared dust-excesses caused from hot dust in their vicinity, with one formation mechanism to form the dust being through the density enhancements from colliding winds in a binary system.

The most famous of these dust-producing systems is WR140 (HD 193793), a binary system containing a WC7 primary and O5.5I-type secondary. It was first discovered to have dust condensation by observation of rapid, periodic brightening in the IR by \cite{Williams,1990MNRAS.243..662W, 2009MNRAS.395.1749W} and is one of two WC binaries with an established visual and double-lined spectroscopic orbit \citep{Thomas}, but the only such system with dust emission. This system has been shown to have a 7.93 year period and a high eccentricity of 0.8993 with masses derived to better than 4$\%$ accuracy \citep{Thomas}. Because of the well-established visual and double-lined orbit, combined with the brightness of the system, WR140's dusty environment has become a fantastic laboratory with which to test dust production models of binary systems and pinpoint the location of the dust production.

\citet{Marchenko} photometrically observed WR140 in the $UBV$ bands as it went through periastron passages in 1993 and 2001. Surprisingly, this is perhaps the only photometric variability study of the system we could find. They found that while in 1993 there were no measurable changes across periastron, the periastron passage in 2001 exhibited a series of dips, which were interpreted as dust eclipses. The $B-V$ versus $U-B$ color-color plot implied that outside of eclipse the star follows the appropriate interstellar reddening line. \citep{Marchenko}.


\section{Observations} \label{sec:style}

New photometric data were obtained following the late-2016 periastron passage of WR140 by the observers of the American Association of Variable Star Observers. These data were taken regularly in the $B$, $V$, and $R$ filters, or at similar wavelengths from DSLR photography. We did not use the $R$-band data in our final analysis as \citet{Marchenko} didn't collect $R$-band data and thus we couldn't make direct comparisons. We found that in all optical bands, there was a single, large dip centered near phase 0.03. There was some subtle evidence of a second noticeable dip following the first one, but the data errors prevent us from stating this with certainty. For the 1993, 2001, and the more recent 2016 periastron passages, we also calculate the color index $B-V$. When multiple telescopes or observers were used, we ensured the data were intercomparable by applying offsets to data that appear based on calibration errors, as was done by \citet{Marchenko}.

\section{Results}

\begin{figure}
    \centering
    \includegraphics[width=\columnwidth]{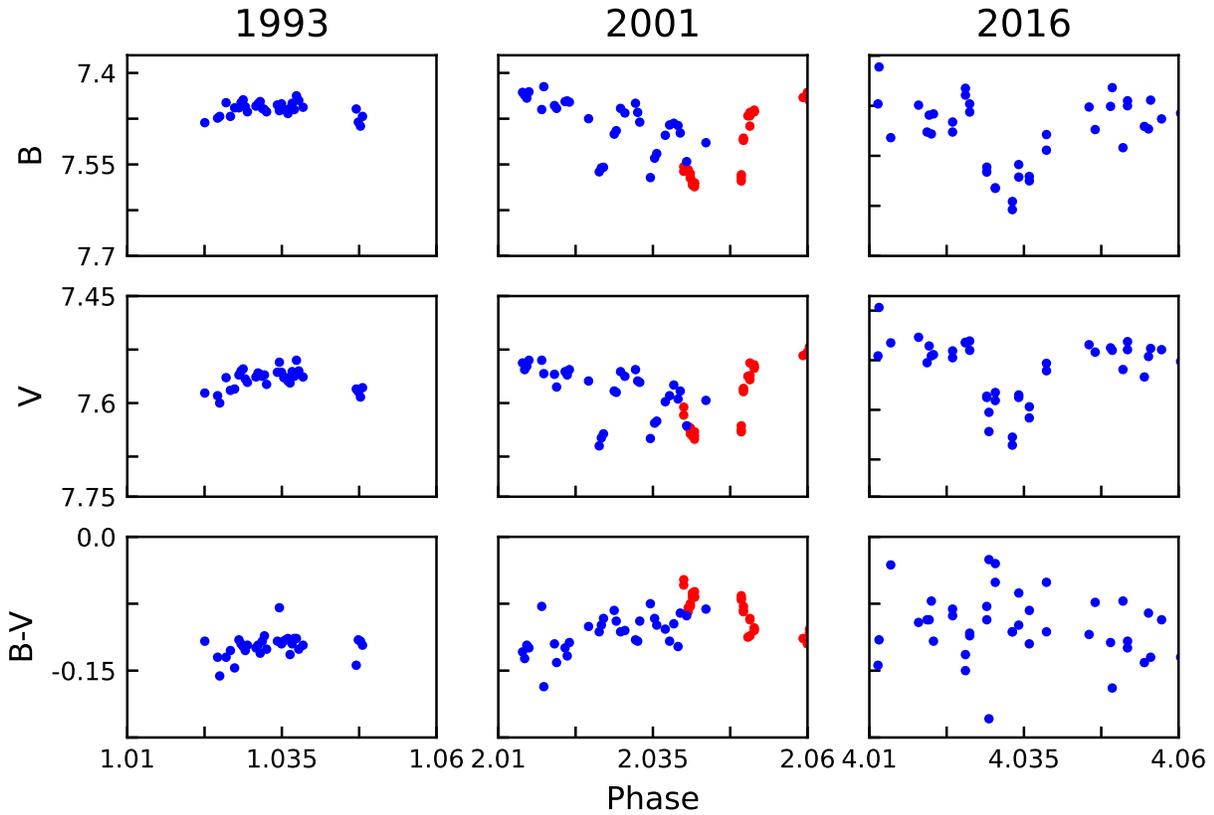}
    \caption{Our final results, using \cite{Marchenko}'s data from 1993 and 2001 as well as data from AAVSO for the 2016 periastron. We completed our analysis in both the $V$- and $B$-band and also show the $B-V$ color for each periastron passage. The different colored points represent different observatories for 1993 and 2001, whereas all of the 2016 data from the AAVSO are in the same blue color. For 1993 and 2001, the blue points represents APT photometry and the red points are from the Rozhen Observatory. There was a slight calibration error that needed to be accounted for when comparing the data, which was originally described by \citet{Marchenko}.}
    \label{fig:my_label}
\end{figure}

In Fig.~\ref{fig:my_label}, we compare the time-series observed during the two periastron passages observed by \citet{Marchenko} and the more recent periastron passage observed by the AAVSO observers. All light curves are plotted using the binary ephemeris from \citet{Thomas}. The 2001 and 2016 periastron passages show variability in deep dips of nearly 0.1 mag in both $B$ and $V$-bands. The dips in 2001 lasted for about 0.03 in phase ($\sim 85$ d), while the singular dip in 2016 lasted for about 0.15 in phase ($\sim 45 d$). In comparison to these events, the earlier 1993 eclipse shows no discernible dip in the optical light curve. We see the system moving to a slightly redder $B-V$ ($\Delta B-V \approx 0.02$) color during these dips in 2001 and 2016, although this is similar to the noise level in 2016. \citet{Marchenko} interpreted the dip in 2001 as being caused by a portion of the dust cloud in our line of sight attenuating the system.

Recently, \citet{2020ApJ...900..190L} developed a geometric model for the colliding winds for a similar WC+O binary, WR112. We utilized the same model but adjusted the orbital parameters for those of the WR140 binary as derived by \citet{Thomas}. Our analysis shows that the dust in our line of sight could be explained is likely near the edge of the projected shock cone. Randomly formed clumps in the region of dust formation can lead to the differences in the observed light curves of the system. 

We note that the orbit of \citet{Thomas} shows some deviations during the most recent 2016 periastron passage compared to the previous periastron passages. Radial velocity differences near periastron, and astrometric differences near apastron probably indicates that the system has a slight precession. Therefore, we examined if changes in the orbital geometry could drastically effect the dust cloud geometry in our line of sight. However, our geometric model shows no major changes in the geometry for the line of sight dust for any value within $\triangle \omega = 10^\circ$, and this favors the idea of random clumps within the line of sight causing the observed differences in the light curves for the three well-observed periastron passages.

In conclusion, we postulate that the photometric dips observed in the WR140 system after a periastron passage is due to dust formed somewhat near an edge of the shocked region. The variations in the dips from orbit to the next likely depend on the localized clumping in this shocked region, whereas the overall amount of dust produced every orbit must remain nearly constant based on the repeatability of the infrared light curve. The understanding of these dust dips could aid in the interpretation of the upcoming, approved {\it JWST}-Early Release Science observations of the dust surrounding the system.

\begin{acknowledgments}

We thank Michael Corcoran, Ryan Lau, Anthony Moffat, Gerd Weigelt, and Peredur Williams for comments that improved this research note.
We acknowledge with thanks the variable star observations from the AAVSO International Database contributed by observers worldwide and used in this research. M.J.P. was funded in part by the Embry-Riddle Aeronautical University's Undergraduate Research Institute.

\end{acknowledgments}

\bibliography{sample631}{}
\bibliographystyle{aasjournal}

\end{document}